\documentclass[a4paper,12pt]{article}
\usepackage{graphicx}
\usepackage{amsmath}
\usepackage{a4}

\input{tcilatex}

\begin{document}

\title{On QCD and $\omega -\rho $ mixing}
\author{N. F. Nasrallah\thanks{%
e-mail: nsrallh@cyberia.net.lb} \\
Faculty of Science, Lebanese University\\
Tripoli, Lebanon.}
\date{}
\maketitle

\begin{abstract}
The evaluation of sum rules and vacuum condensates used in the calculation
of the mixing parameters of the $\omega -\rho $ system involve dangerous
cancellations of large numbers. In this note the contribution of the
continuum is eliminated in order to avoid such a large source of
uncertainty. It is also seen that the hypothesis of vacuum saturation used
in the estimate of the dimension 6 condensate is inadequate. The value
obtained for the parameter $\beta $ which governs the off-shell variation of
the $\omega -\rho $ mixing is $\beta $=.25 .
\end{abstract}

Since Shifman,Vainshtein and Zakharov \cite{c1} first applied QCD sum rules
to the problem of $\omega $-$\rho $ mixing in order to probe the mechanism
of isotopic symmetry breaking several attempts have been made to obtain the $%
\omega $-$\rho $ mixing parameters in terms of the quark masses and the
various quark and gluon field condensates \cite{c2}, \cite{c3}, \cite{c4}.
The values obtained for the mixing parameters vary widely in the literature.
The reason \ for this instability is easily found: the sum rules for the
problems are integrals over the difference of two nearly equal
contributions, the one of the 2$\pi $ intermediate states dominated by the $%
\rho -\mathrm{meson}$ and the one of the 3$\pi $ intermediate states
dominated by the $\omega $-meson. The integrals being the difference of
large numbers it is not surprising that the various approximations used,
which do not necessarily evenly effect each contribution, can result in
large perturbations in the values obtained. The same applies to the estimate
of the vacuum condensates of the operator product expression which are
themselves differences of nearly equal numbers and for the evaluation of
which it is certainly overoptimistic to trust the vacuum saturation
hypothesis as has been done in the literature. In this note no attempt will
be made to estimate the contribution of the continuum, this contribution
will rather be eliminated by an appropriate choice of a damping factor in
the sum rules. It will also be shown that the vacuum saturation hypothesis
must necessarily be relaxed in the determination of the dimension-6
condensate which will treated as an unknown.

The correlator relevant to the problem is:

\begin{equation}
\Pi _{\mu \nu }=i\int \exp {iqx}\left\langle 0\mid \mathrm{TV}_{\mu
}^{(1)}(x)\mathrm{V}_{\mu }^{(0)}(0)\mid 0\right\rangle dx
\end{equation}

Where the superscripts refer to isospin

$\ \ \ \ \ \ \ \ \ \ \ \ \ \ \ \ \ \ \ \ \ \ \ \ \ \ \ \ \ \ \ \ \ \ \ \ \ \
\ \ \ \ \ \ \ \ \ \ \ \ \ \ \ \ \ \ \ \ \ $%
\begin{equation}
\ \ \ \ \ \ \ \ \ \ \ \ \mathrm{V}_{\mu }^{(1)}=\frac{1}{2}(\bar{u}\gamma
_{\mu }u-\bar{d}\gamma _{\mu }d)\text{ \ \ \ \ \ , \ \ \ \ \ }\mathrm{V}%
_{\nu }^{(0)}=\frac{1}{2}(\bar{u}\gamma _{\mu }u+\bar{d}\gamma _{\mu }d)
\end{equation}
$\ \ \ \ \ $

and from current conservation

\begin{equation}
\Pi _{\mu \nu }(t=q^{2})=(q_{\mu }q_{\nu }-q^{2}g_{\mu \nu })\Pi (t)
\end{equation}

$\Pi (t)$ is an analytic function in the complex t-plane with a cut along
the positive $t$ axis starting at $t=4m_{\pi }^{2}$. Furthermore the
asymptotic behaviour of $\Pi (t)$ in the deep Euclidean region is obtained
from QCD \cite{c1}:

\begin{equation}
12\Pi ^{\mathrm{QCD}}(t)=c_{0}\ln (-t)+\frac{c_{1}}{t}+\frac{c_{2}}{t^{2}}+%
\frac{c_{3}}{t^{3}}+\cdots
\end{equation}

with

\begin{eqnarray}
c_{0} &=&-\frac{\alpha _{em}}{16\pi ^{2}}  \notag \\
c_{1} &=&-\frac{3}{2\pi ^{2}}(m_{d}^{2}-m_{u}^{2})\simeq 0  \notag \\
c_{2} &=&2f_{\pi }^{\,2}m_{\pi }^{2}\left( \frac{m_{d}-m_{u}}{m_{d}+m_{u}}%
\right) \left[ 1+\frac{\gamma }{2+\gamma }\left( \frac{m_{d}+m_{u}}{%
m_{d}-m_{u}}\right) \right]  \notag \\
c_{3} &=&\frac{\pi \alpha _{s}}{6}\langle 0\mid (\bar{u}\gamma _{\alpha
}\gamma _{5}\lambda ^{\alpha }u)^{2}-(\bar{d}\gamma _{\alpha }\gamma
_{5}\lambda ^{\alpha }d)^{2}\mid 0\rangle \smallskip  \notag \\
&&+\frac{\pi \alpha _{s}}{27}\langle 0\mid (\bar{u}\gamma _{\alpha }\lambda
^{\alpha }u)^{2}-(\bar{d}\gamma _{\alpha }\lambda ^{\alpha }d)^{2}\mid
0\rangle \\
&&+\frac{2\pi \alpha _{em}}{3}\langle 0\mid \frac{4}{9}(\bar{u}\gamma
_{\alpha }\gamma _{5}\lambda ^{\alpha }u)^{2}-\frac{1}{2}(\bar{d}\gamma
_{\alpha }\gamma _{5}\lambda ^{\alpha }d)^{2}\mid 0\rangle  \notag \\
&&+\frac{4\pi \alpha _{em}}{27}\langle 0\mid \frac{4}{9}(\bar{u}\gamma
_{\alpha }u)^{2}-\frac{1}{9}(\bar{d}\gamma _{\alpha }d)^{2}\mid 0\rangle 
\notag \\
\gamma &\equiv &\frac{\langle 0\mid \bar{d}d\mid 0\rangle }{\langle 0\mid 
\bar{u}u\mid 0\rangle }-1  \notag
\end{eqnarray}
\qquad

In the expression for $c_{2}$ use has been made of the
Gell-Mann-Oakes-Renner relation $(m_{n}+m_{d})\,\langle 0\mid \bar{u}u+\bar{d%
}d\mid 0\rangle =-2f_{\pi }^{2}m_{\pi }^{2}$.

In the expression for $c_{3}$, the gluon exchange and the photon exchange
contribution are easily recognizable, the former dominant ones are clearly
differences of nearly equal numbers.

In the narrow width approximation, the absorptive part of the hadronic
amplitude is

\begin{equation}
\frac{12}{\pi }\func{Im}\Pi (t)=f_{\rho }\delta (t-m_{\rho }^{2})-f_{\omega
}\delta (t-m_{\omega }^{2})+f_{\phi }\delta (t-m_{\phi }^{2})+\mathrm{%
continuum}
\end{equation}

where the continuum is approximated by broad $\rho ^{\prime }$ and $\omega
^{\prime }$ resonances. Instead of trying to evaluate the contribution of
the latter which, as discussed above, involve large cancellations it is
preferable to eliminate this contribution. To this effect consider the
integral in the complex $t$-plane over the closed contour consisting of a
circle of large radius $R$ and two straight lines immediately above and
below the cut which run from threshold to $R$, this integral vanishes by
virtue of Cauchy's theorem 
\begin{equation}
\frac{12}{\pi }\int_{th}^{R}\func{Im}\Pi (t)(t-m^{\prime 2})^{2}dt=-\frac{12%
}{2\pi i}\oint \Pi (t)(t-m^{\prime 2})^{2}dt  \label{7}
\end{equation}

where $m^{\prime 2}=(m_{\rho }^{\prime 2}+m_{\omega }^{\prime 2})/2$. The
factor $(t-m^{\prime 2})^{2}$ practically eliminates the contribution of the
resonances. Moreover it will not introduce condensates of dimension higher
than 6 in the calculation. The contribution of the $\phi $ can easily be
estimated using 
\begin{equation}
f_{\phi }\simeq \delta ^{2}\left[ f_{\omega }+\frac{(f_{\omega }-f_{\rho })}{%
\delta m^{2}}(m_{\phi }^{2}-m^{2})\right] 
\end{equation}

where $m^{2}=(m_{\rho }^{2}+m_{\omega }^{2})/2$ , $\delta m^{2}=(m_{\omega
}^{2}-m_{\rho }^{2})$

and where $\delta =.065$ rad is the deviation of the octet singlet mixing
angle from its ideal value $(\phi =\overline{s}s)$.

The contribution of the $\phi $ turns out to be small even when compared to
the difference of the $\rho $ and $\omega $ contributions.

In the integral over the circle of radius $R$ appearing on the r.h.s of Eq. (%
\ref{7}) $\Pi $ is well approximated by $\Pi ^{\mathrm{QCD}}$ except
possibly for a small region near the real axis. The narrow resonance
approximation is certainly not adequate for the contribution of the $\rho -%
\mathrm{meson}$. In order to correct for this we represent the $\rho $ by a
Breit-Wigner of finite extent normalized to unity, i.e.

\begin{equation}
f_{\rho \,}\delta (t-m_{\rho }^{2})\longrightarrow \frac{m_{\rho }\Gamma
_{\rho }}{2\arctan \left( \frac{w}{m_{\rho }\Gamma _{\rho }}\right) }\frac{%
f_{\rho }(t)}{(t-m_{\rho }^{2})^{2}+m_{\rho }^{2}\Gamma _{\rho }^{2}}
\end{equation}

for $m_{\rho }^{2}-w\leq t\leq m_{\rho }^{2}+w\ \ \ \ \ ,\ \ \ \ \ \ \ \ \ \
w=.4\mathrm{GeV}^{2}$

and

\begin{equation}
f_{\rho }(t)\simeq f_{\rho }+\frac{(f_{\omega }-f_{\rho })}{\delta m^{2}}%
(t-m_{\rho }^{2})
\end{equation}

Defining the mixing parameters $\beta $ and $\xi $

\begin{equation}
(f_{\rho }+f_{\omega })=\frac{2m^{4}}{\delta m^{2}}\xi \text{ \ \ \ \ , \ \
\ \ \ \ }(f_{\omega }-f_{\rho })=m^{2}\beta \xi
\end{equation}

and performing the integrals yields

\begin{eqnarray}
&&\beta \xi \left[ m^{2}\frac{(m^{\prime 2}-m^{2})}{m^{\prime 4}}+\frac{1}{2}%
m^{2}\delta _{1}+\frac{m^{4}}{\delta m^{2}}\delta _{2}\right] -\xi \left[
2m^{4}\frac{\left( m^{\prime 2}-m^{2}\right) }{m^{\prime 4}}+\frac{m^{4}}{%
\delta m^{2}}\delta _{1}\right]   \notag \\
&=&c_{0}\left[ \frac{1}{3}\frac{R^{3}}{m^{\prime 4}}-\frac{R^{2}}{m^{\prime
2}}-R\right] -\frac{2c_{2}}{m^{\prime 2}}+\frac{c_{3}}{m^{\prime 4}}
\label{12}
\end{eqnarray}
where 
\begin{eqnarray}
\delta _{1} &=&\frac{1}{m^{\prime 4}}\left( \frac{m\Gamma _{\rho }^{2}w}{%
\arctan \left( \frac{w}{m_{\rho }\Gamma _{\rho }}\right) }-m^{2}\Gamma
_{\rho }^{2}\right)   \notag \\
\delta _{2} &=&2\frac{\left( m^{\prime 2}-m^{2}\right) }{m^{2}}\delta _{1}
\end{eqnarray}

$\delta _{1}$ and $\delta _{2}$ both vanish in the narrow width
approximation. Because $c_{0}$ is \ small the dependance on $R$ is very
weak. $\xi $ is extracted from electroproduction experiments \cite{c2} $\xi
=(1.13\pm .13)\times 10^{-3}$. If \ we use the standard value $\frac{%
m_{d}-m_{u}}{m_{d}+m_{u}}=.29$ to evaluate $c_{2}$ and if we use the value
of $c_{3}$ resulting from the vacuum saturation hypothesis 
\begin{equation}
c_{3}=\frac{448\pi }{81}\alpha _{s}\left| \langle \bar{q}q\rangle \right|
^{2}\left[ \frac{\alpha _{m}}{8\alpha _{s}\left( \mu ^{2}\right) }-\gamma %
\right] \simeq .18\times 10^{-4}\mathrm{GeV}^{6}
\end{equation}

with 
\begin{equation*}
R\simeq 2.5-3\mathrm{GeV}^{2}\ \ ,\ \ \gamma =-.01
\end{equation*}

we get 
\begin{equation}
\beta =.46
\end{equation}

close to $\beta =.50$ used by SVZ.

It is instructive to proceed further and to use the first moment integral

\begin{equation}
\frac{1}{\pi }\int_{th}^{R}t\left( t-m^{\prime 2}\right) ^{2}\func{Im}\Pi
(t)=\frac{1}{2\pi i}\oint \Pi (t)t\left( t-m^{\prime 2}\right) ^{2}dt
\label{16}
\end{equation}

The price to pay consists in the introduction of an unknown condensate $c_{4}
$ in the problem whose contribution is however fortunately strongly damped,
Eq. (\ref{16}) yields then an additional relation

\begin{eqnarray}
&&\beta \xi \left[ m^{4}\frac{\left( m^{\prime 2}-m^{2}\right) ^{2}}{%
m^{\prime 4}}+\frac{1}{2}m^{4}\Delta _{1}-\frac{m^{6}}{\delta m^{2}}\Delta
_{2}\right]  \notag \\
&&+\xi \left[ \frac{m^{4}}{m^{\prime 4}}\left( m^{\prime 2}-m^{2}\right)
\left( m^{\prime 2}-3m^{2}\right) -\frac{m^{6}}{8m^{2}}\Delta _{1}\right] 
\notag \\
&=&c_{0}\left[ \frac{1}{4}\frac{R^{4}}{m^{\prime 4}}-\frac{2}{3}\frac{R^{3}}{%
m^{\prime 2}}+\frac{1}{2}R^{2}\right] +c_{2}-2\frac{c_{3}}{m^{\prime 2}}+%
\frac{c_{4}}{m^{\prime 4}}  \label{17}
\end{eqnarray}

with 
\begin{eqnarray}
\Delta _{1} &=&-\frac{\left( 2m^{\prime 2}-3m^{2}\right) }{m^{2}}\delta _{1}
\\
\Delta _{2} &=&\frac{1}{3\arctan \left( \frac{w}{m_{\rho }\Gamma _{\rho }}%
\right) }\left( \frac{\Gamma _{\rho }}{m}\right) \frac{w^{3}}{m^{2}m^{\prime
4}}-\left( \frac{\Gamma _{\rho }}{m}\right) ^{2}\left( \delta
_{1}+2m^{2}\Gamma _{\rho }^{2}\right)   \notag \\
&&+\frac{\left( m^{\prime 2}-m^{2}\right) \left( m^{\prime 2}-3m^{2}\right) 
}{m^{4}}\delta _{1}  \notag
\end{eqnarray}

which vanish in the narrow width approximation.

If $c_{4}/c_{3}\simeq 0.6$\textrm{GeV}$^{2}-1$\textrm{GeV}$^{2}$ as
expected, the contribution of the last term on the r.h.s of Eq. (\ref{17})
amounts to no more than 15-20\% of the term preceding it, we shall use this
term only to estimate the error.

If we use the value $c_{3}=.18\times 10^{-4}\mathrm{GeV}^{6}$ given by the
vacuum saturation hypothesis we get from Eq. (\ref{17}) and Eq. (\ref{12})

\begin{equation}
\xi =0.50\times 10^{-3}\text{ \ \ \ \ \ \ \ \ \ \ , \ \ \ \ \ \ \ \ \ \ \ \
\ }\beta =.11
\end{equation}

The disagreement of the calculated value of $\xi $ with the value extracted
from experiment is a clear indication that vacuum saturation fails to yield
the correct value for $c_{3}$.

If we treat $c_{3}$ and $\beta $ as unknowns and insert the experimental
value of $\xi $ in Eq. (\ref{17}) and Eq. (\ref{12}) we get instead

\begin{equation}
\beta =.25\text{ \ \ \ \ \ \ \ \ \ \ , \ \ \ \ \ \ \ \ \ \ \ \ \ \ \ }%
c_{3}=-1.93\times 10^{-4}\mathrm{GeV}^{6}
\end{equation}

The error on the values above is estimated to be of the order of $20\%$.

The fact that the value of the dimension 6 condensate deviates from the
vacuum saturation value has already been noted elsewhere \cite{c5}, \cite{c6}%
, this effect is here amplified by the fact that we are dealing with the
difference of large numbers. The parameter $\beta $ which governs the
off-shell variation of the $\rho -\omega $ mixing is relevant in the study
of the charge independence of the nuclear force in particular in the
explanation of the Okamoto-Nolen-Schiffer anomaly \cite{c7}

It is finally worth examining the scenario known as generalized Chiral
perturbation theorems $G\chi pt$ \cite{c8} in which the condensate $\langle 
\bar{q}q\rangle $ is much smaller than in the standard case, the coefficient 
$c_{2}$ becomes in this case negligible and the corresponding calculated
values of $\beta $ and $c_{3}$ are then

\begin{equation}
\beta =.33\text{ \ \ \ \ \ \ \ \ \ \ \ \ , \ \ \ \ \ \ \ \ \ \ \ \ \ }%
c_{3}=-3.5\times 10^{-4}\mathrm{GeV}^{6}
\end{equation}

\bigskip

\bigskip

\bigskip

\textbf{Acknowledgments:} I thank the Academie de Versailles for its
generous support, the IPN, Orsay, where the essential part of this work was
done, for their kind hospitality as well as Bachir Moussallam and Hagop
Sazdjian for useful discussions.


\newpage

\begin{thebibliography}{99}
\bibitem{c1}  \label{a1}M.A. Shifman, A. I. Vainshtein, V.I. Zakharov, Nucl.
Phys. \textbf{B147 }(1979) 519.

\bibitem{c2}  \label{a2}T. Hatsuda, E. M. Henley, Th. Meissner, G. Krein,
Phys. Rev. \textbf{C49 }(1994) 452 .

\bibitem{c3}  \label{a3}K. Maltman, Phys. Rev. \textbf{D53 }(1996) 2563 .

\bibitem{c4}  \label{a4}M. I. Iqbal, X. Jin, D. Leinweber, Phys. Lett. 
\textbf{B386 }(1996) 55 .

\bibitem{c5}  \label{a5}V. Gimenez, V. Bordes, J. Pennarocha, Nucl. Phys. 
\textbf{B357 }(1991) 3 .

\bibitem{c6}  \label{a6}M. Davier, L. Girlanda, A. H\"{o}cker, J. Stern,
Phys. Rev. \textbf{D58 }(1998) 096014.

\bibitem{c7}  \label{a7}J. A. Nolen, Jr. and J. P. Schiffer, Annu. Rev.
Nucl. Sci. \textbf{19} (1969) 471.

S.Shlomo, Rep. Prog. Phys. \textbf{41 }(1978) 957.

\bibitem{c8}  \label{a8}N. H. Fuchs, H. Sazdjian, J. Stern, Phys. Lett. 
\textbf{B269} (1991) 183.

J. Stern, H. Sazdjian, N. H. Fuchs, Phys. Rev. \textbf{D47} (1993) 3814.
\end{thebibliography}
\end{document}